\documentclass[aps,pre,twocolumn,superscriptaddress,showpacs]{revtex4-1}
\usepackage{graphicx}
\usepackage{bm}
\usepackage{hyperref}
\usepackage[mathlines]{lineno}
\usepackage{amsmath}
\usepackage{amssymb}
\usepackage{times}
\renewcommand{\d}{\mathrm{d}}

\begin{document}

\title{Deterministic fractals: extracting additional information\\
from small-angle scattering data}

\author{A. Yu. Cherny}
\email[]{cherny@theor.jinr.ru}
\affiliation{Joint Institute for Nuclear Research, Dubna
141980, Moscow region, Russian Federation}

\author{E. M. Anitas}
\email[]{anitas@theor.jinr.ru}
\affiliation{Joint Institute for Nuclear Research, Dubna
141980, Moscow region, Russian Federation} \affiliation{Horia Hulubei National Institute
of Physics and Nuclear Engineering, RO-077125 Bucharest-Magurele, Romania}

\author{V. A. Osipov}
\email[]{osipov@theor.jinr.ru}
\affiliation{Joint Institute for Nuclear Research, Dubna
141980, Moscow region, Russian Federation}

\author{A. I. Kuklin}
\email[]{kuklin@nf.jinr.ru} \affiliation{Joint Institute for Nuclear Research, Dubna
141980, Moscow region, Russian Federation}

\date{\today}

\begin{abstract}
The small-angle scattering curves of deterministic mass fractals are studied and 
analyzed in the momentum space. In the fractal region, the curve $I(q)q^D$ is found to 
be log-periodic with a good accuracy, and the period is equal to the scaling factor of 
the fractal. Here  $D$ and $I(q)$ are the fractal dimension and the scattering intensity, 
respectively. The number of periods of this curve coincides with the number of fractal 
iterations. We show that the log-periodicity of $I(q)q^D$ in the momentum space is 
related to the log-periodicity of the quantity $g(r)r^{3-D}$ in the real space, where 
$g(r)$ is the pair distribution function. The minima and maxima positions of the 
scattering intensity are estimated explicitly by relating them to the pair distance 
distribution in the real space. It is shown that the minima and maxima are damped with 
increasing polydispersity of the fractal sets; however, they remain quite pronounced 
even at sufficiently large values of polydispersity. A generalized self-similar Vicsek 
fractal with controllable fractal dimension is introduced, and its scattering properties 
are studied to illustrate the above findings. In contrast with the usual methods, the 
present analysis allows us to obtain not only the fractal dimension and the edges of 
the fractal region, but also the fractal iteration number, the scaling factor, and the number of 
structural units from which the fractal is composed.
\end{abstract}


\maketitle

\section{\label{sec:intro}{Introduction}}

Modern experimental techniques in materials science for preparing hierarchically organized systems at nano and micro scales open up new possibilities in controlling their functions and properties. Considerable scientific and technological efforts have been directed to the development of such systems \cite{harald94, lomander05, bystrova05, mayama06, newkome06, barth07, wang07, cerofolini08, yamazaki09,  polshettiwar09, stoliar10}. The hierarchical structures, like fractals \cite{mandelbrot82} or biological objects \cite{lebedev08}, manifest themselves in the electromagnetic \cite{huang09book}, statistical \cite{luscombe85}, dynamical \cite{galiceanu10} or optical \cite{pustovit02, allain86} properties, and therefore one of the basic challenges is to understand correlations between these properties and material's microstructure \cite{pynn85, nakayama03}.

Experimentally, the material's microstructure can be determined by means of scattering 
techniques, which yield elastic cross section per unit solid angle as a function of 
momentum transfer. The cross section, usually normalized per unit volume of a sample, is 
called scattering intensity: $I(q)\equiv \displaystyle{\frac{1}{V'}\frac{\d \sigma}{\d 
\Omega}}$. The scattering angle $2\vartheta$ is related to the momentum transfer $\hbar 
q$ by equation $q=4 \pi\lambda^{-1}\sin\vartheta$, where $\lambda$ is the radiation 
wavelength. The typical range in which the materials exhibit a hierarchical structure is 
within $d=1 \div 1000~\mathrm{nm}$ (see, e.g., Ref.~\cite{lebedev05}). The structural 
characteristics in this range are probed well when the scattering wave vector lies 
within $q\simeq 2\pi/d=10^{-3} \div 1~\text{\AA}^{-1}$, which corresponds to 
\emph{small-angle} X-ray or neutron scattering (SAXS/SANS) \cite{gf55, fs87, lz02}. This 
important investigation technique has become a powerful tool for studying the fractal 
microstructure (see reviews \cite{mh87, teixeira88, schmidt91, schmidt91book} and 
references therein).

Presently, various deterministic fractals can be artificially created \cite{mayama06, 
newkome06, barth07, cerofolini08, polshettiwar09} due to a rapid progress in 
nanotechnologies. This makes it relevant and promising to study the connections between 
their scattering properties and their microstructure. Deterministic fractals frequently allow 
an analytical description of small-angle scattering (SAS) and thus give us ``exactly 
solvable models", very useful for understanding the scattering properties of fractals. 
By choosing parameters of deterministic fractals at random (say, by introducing 
polydispersity), one can understand the basic properties of random fractals as well 
\cite{mandelbrot84, kjems85:book}.

An essential feature of fractal is Hausdorff (fractal) dimension. While it can be rigorously defined (see Appendix \ref{sec:hd}), it is convenient in practice to adopt a simple descriptive definition of the dimension $D$ \cite{mandelbrot82, pfeifer83, barnsley88, gouyet96:book}: $N\varpropto (1/a)^D$ for $a\to 0$, where $N$ is the minimum number of open sets of diameter $a$ needed to cover a fractal. Non-random (deterministic) fractals are generated by deterministic processes, in particular, by iterative rules. This implies the presence of an initial set (initiator) and a generator (iterative operation). The number of iterative operations is called fractal iteration. Let us consider a finite iteration of mass fractal, consisting of simple units, say, balls of radius $a$.  Let the fractal size be $l$. If $N$ is the total number of balls then the fractal dimension $D$ is defined by the asymptotics
\begin{equation}
N \varpropto (l/a)^{D} \label{eq:fd}
\end{equation}
for a large number of iterations, which assumes $a \to 0$ (from the physical point of view, $a \ll l$). One can use this definition for a given finite iteration ($a=\mathrm{const}$) in order to estimate the number of balls enclosed by an imaginary sphere of radius $r$ with a ball in the center. Equation (\ref{eq:fd}) yields $N(r) \varpropto (r/a)^{D} \varpropto r^{D}$. This estimation is valid within the \emph{fractal region} $l_\mathrm{min}\lesssim r \lesssim l$, in which fractal properties can be observed experimentally. Here $l_\mathrm{min}$ is the minimal distance between the ball centers.

A main indicator of the fractal structure is the power-law dependence between the small-angle scattering intensity and the absolute value of scattering vector \cite{martin85, kjems85a:book, martin86, freltoft86}. For a mass-fractal, it takes the form
\begin{equation}
I(q) \varpropto q^{-D}. \label{intensityreduced}
\end{equation}
The fractal region in the real space implies the fractal region in the reciprocal space
\begin{equation}
1/l \lesssim q \lesssim 1/l_\mathrm{min},
\label{fractalrangegeneral}
\end{equation}
for which Eq.~(\ref{intensityreduced}) is applicable. Equation (\ref{intensityreduced}) is quite general, because the fractal dimension essentially regulates the spatial correlations between fractal units. Indeed, the scattering intensity is proportional to the Fourier transform of the pair distribution function $g(\bm{r})$, describing the spatial correlations between particles inside a fractal (see Sec.~\ref{sec:gendef} below). In more detail, once a particle lies in the coordinate origin then the number of other particles in the volume $\d^3 r$ near the radius vector $\bm{r}$ is equal to $\d N=n g(\bm{r})\d^3 r$, where $n$ is the average particle density. If the pair distribution function is radially symmetric, we obtain $N(r)=\int_{0}^{r} n g(r') 4\pi r'^2\d r'$ for the total number of particles in the sphere of radius $r$. On the other hand, the fractal dimension implies $N(r)\varpropto r^D$, as discussed above. It follows that
\begin{equation}
g(r)\varpropto r^{D-3}
\label{grpower}
\end{equation}
for $r\lesssim l$ \cite{forrest79, witten81}, which leads to Eq.~(\ref{intensityreduced}) in accordance with Erd\'elyi's theorem for asymptotic expansion of Fourier integrals \cite{erdelyi56:book}.

Thus, three parameters can be extracted from experimental fractal scattering intensities: the exponent and the edges of the fractal region in the $q$-space, which appear as ``knees" in the scattering line on a logarithmic scale. Other parameters are not usually obtained from SAS curves. For random fractals, it is rather difficult to extract more information from the scattering, because a fine structure of particle correlations is usually smeared due to the randomness. One can expect that deterministic fractals, being more ordered, allow us to obtain additional information from the scattering data. In this paper we suggest a scheme for estimating the iteration number, the scaling factor and the total number of structural units in deterministic mass fractals from the scattering curves.

The present analysis of deterministic fractals is based on the method of calculating the 
scattering amplitude, developed in the papers 
\cite{kjems85:book,allain85:book,allain86,sx86} (for the generalization, see 
Ref.~\cite{lidar96}). Recently, analytical calculations of SAS intensity from the 3D 
triadic Cantor and Vicsek \cite{vicsek83} sets were reported in Ref.~\cite{cherny110}. 
In the previous paper \cite{cherny210}, the generalized three-dimensional (3D) Cantor set was suggested, 
whose dimension can vary from 0 to 3 by means of the scaling factor. If the SAS 
is considered from monodispersive sets, which are randomly oriented and placed, then the 
scattering intensities represent minima and maxima superimposed on a power-law decay, 
with the exponent equal to the fractal dimension of mass fractal. This dependence of the 
intensity on momentum is called the generalized power-law decay. As was shown 
\cite{sx86,cherny110,cherny210}, the minima and maxima are damped with increasing 
polydispersity of the fractal sets. The physical reasons for such behavior are quite clear: 
the fractal dimension dictates the power law (\ref{grpower}) for the particle 
correlations \emph{only on the average}. In this paper, we show that in a deterministic 
fractal, the pair distribution function $g(r)$ also obeys the generalized power law [see 
Eq.~(\ref{eq:grexp}) below], that is, it exhibits minima and maxima on the power-law 
decay (\ref{grpower}). This structure appears due to clusterization and 
intimately relates to the fractal scaling factor. Polydispersity smears the spatial 
distribution between fractal units. Strongly developed polydispersity thus leads to the 
simple power-law behavior (\ref{intensityreduced}) and (\ref{grpower}). In this paper, 
we construct and consider the generalized self-similar Vicsek fractals (GSSVF) as an 
example in order to elucidate the above scattering properties.

The paper is organized as follows: in Sec.~\ref{sec:theory} we emphasize some important issues concerning SAS. Section \ref{sec:construction} describes the construction of GSSVF with controllable dimension, governed by the scaling factor. In the subsequent section, we derive analytically the scattering amplitude for GSSVF and calculate its scattering properties: the intensity, structure factor, and radius of gyration. The influence of polydispersity on the fractal scattering properties is considered in Sec.~\ref{sec:polydisperse}. The spatial correlation functions of the fractal are obtained and interpreted in Sec.~\ref{sec:geometry}. In Conclusion we discuss the obtained results and promising prospects of the developed analysis.

\section{\label{sec:theory}{General remarks on small-angle scattering}}

Let us consider the SAS scattering (neutron, X-ray, light, or electron diffraction) on a sample consisting of microscopic objects with the scattering length $b_j$. If the multiscattering processes are neglected (which is a very good approximation usually), then the differential cross section of the sample is given by \cite{fs87} $\d\sigma/\d\Omega=|A(\bm{q})|^2$, where $A(\bm{q})\equiv \int_{V'} \rho_\mathrm{s}(\bm{r}) e^{i \bm{q}\cdot\bm{r}}\d^3 r$ is the total scattering amplitude, and $V'$ is the total volume irradiated by the incident beam. The scattering length density can be defined with the help of Dirac's $\delta$-function as $\rho_\mathrm{s}(\bm{r})=\sum_j b_j\delta(\bm{r}-\bm{r}_j)$, where $\bm{r}_j$ are the microscopic object positions.

In this paper, we restrict ourselves to two-phase systems, which are composed of homogeneous units of ``mass" density $\rho_\mathrm{m}$. The units are immersed into a solid matrix of ``pore" density $\rho_\mathrm{p}$. A constant shift of the scattering length density in the overall sample is important only for small values of wave vector $q\lesssim 2\pi/(V')^{1/3}$, which are usually beyond  the resolution of scattering device. Therefore, by subtracting the ``pore" density, we can consider the system as if the units were ``frozen" in a vacuum and had the density $\Delta\rho =\rho_\mathrm{m} - \rho_\mathrm{p}$. The density $\Delta\rho$ is called scattering contrast.

In practice, it is convenient to represent the total scattering amplitude as a sum of amplitudes of rigid objects. For instance, considering the scattering from stiff fractals, whose spatial positions and orientations are uncorrelated, one can choose them as the objects. Then the scattering intensity (that is, the differential cross section per unit volume of the sample) is given by
\begin{equation}
I(q) = n |\Delta\rho|^{2} V^{2}\left\langle \left|F(\bm{q})\right|^{2}\right\rangle,
\label{intensitygeneral}
\end{equation}
where $n$ is the fractal concentration, $V$ is the volume of each fractal, and $F(\bm{q})$ is the normalized formfactor
\begin{equation}
F(\bm{q})=\frac{1}{V}\int_{V}e^{-i\bm{q}\cdot\bm{r}}\mathrm{d}\bm{r},
\label{formfactorgeneral}
\end{equation}
obeying the condition $F(0)=1$. Here, the brackets $\left\langle \cdots \right\rangle$ stand for the ensemble averaging over all orientations of the fractals. If the probability of any orientation is the same, then it can be calculated by averaging over all directions $\bm{n}$ of the momentum transfer $\bm{q}=q \bm{n}$, that is, by integrating over the solid angle in the spherical coordinates ${q}_{x}=q \cos\varphi \sin\vartheta$, ${q}_{y}=q \sin\varphi \sin\vartheta$ and ${q}_{z}=q \cos\vartheta$
\begin{equation}
\langle f(q_x,q_y,q_z)
\rangle=\frac{1}{4\pi}\int_{0}^{\pi}d\vartheta\sin\vartheta\int_{0}^{2\pi}d
\varphi\,f(q,\vartheta,\varphi). \label{aver}
\end{equation}
The intensity in zero angle results directly from Eq.~(\ref{intensitygeneral})
\begin{equation}
I(0) = n |\Delta\rho|^{2} V^{2}. \label{intensityinzerogeneral}
\end{equation}
Given the contrast, concentration of fractals, and the absolute value of intensity, the fractal volume can be determined from Eq.~(\ref{intensityinzerogeneral}).

Once a deterministic fractal is composed of the same objects, say, $N$ balls of the same radius $R$, then the formfactor can be written as
\begin{equation}
F(\bm{q})=\rho_{\bm{q}} F_0(q R)/N. \label{fqsq}
\end{equation}
Here the ball formfactor of unit radius is given by \cite{fs87}
\begin{equation}
F_{0}(z)=3(\sin z - z \cos z)/z^{3}, \label{ballformfactor}
\end{equation}
and $\rho_{\bm{q}}=\sum_{j}e^{-i\bm{q}\cdot\bm{r}_{j}}$ is the Fourier component of the density of ball centers, where $\bm{r}_{j}$ are the center-of-mass positions of balls. By substituting Eq.~(\ref{fqsq}) into Eq.~(\ref{intensitygeneral}), we obtain for the scattering intensity
\begin{equation}
I(q)=I(0)S(q)|F_{0}(q R)|^2/N, \label{intsq}
\end{equation}
where we put by definition (see, e.g., \cite{march67:book})
\begin{equation}
S(q)\equiv\langle\rho_{\bm{q}}\rho_{\bm{-q}}\rangle/N.
\label{sq}
\end{equation}
The quantity $S(q)$ is called structure factor, and it is intimately connected to the pair distribution function (see the discussion in Sec.~\ref{sec:geometry} below). The structure factor carries information about the relative positions of the balls in the fractal and can be rewritten as
\begin{equation}
S(q)=\frac{1}{N}\sum_{j,k=1}^{N}\big\langle\exp[-i\bm{q}\cdot(\bm{r}_{j}-\bm{r}_{k})]\big\rangle.
\label{sqsum}
\end{equation}
It obeys the relations $S(0)=N$ and $S(q)\simeq 1$ for $q\gtrsim 1/l_\mathrm{min}$. The former relation follows from the definition (\ref{sq}), and the latter is fulfilled, because the contribution of non-diagonal terms in the r.h.s. of Eq.~(\ref{sqsum}) $\sum_{j\not=k}\langle e^{-i\bm{q}\cdot(\bm{r}_{j}-\bm{r}_{k})} \rangle/N$ tends to zero in the limit of large momentum due to the randomness of the phase. This limit is analogous to the limit of geometrical optics.

Let us emphasize that although Eq.~(\ref{intsq}) is quite general, the choice of 
structure factor $S(q)$ and the formactor $F_0$ is rather arbitrary. It depends on the 
choice of the scattering units, to which the formfactor $F_0$ is related. This choice 
means just a regrouping of terms in the total scattering amplitude $A(\bm{q})$, 
introduced in the beginning of this section. For example, if we associate $F_0$ with the 
formfactor of entire fractal, then the structure factor describes the spatial 
correlations between different fractals. In this case, the structure factor equals one 
for non-zero momenta, because the fractal positions are assumed to be completely 
uncorrelated. For this choice, Eq.~(\ref{intensitygeneral}) is analogous to 
Eq.~(\ref{intsq}) with $S(q)=1$.

An important characteristic in SAS is radius of gyration $R_\mathrm{g}$. It can be defined from the intensity expansion in the Guinier regime $q\lesssim 1/l_0$ \cite{fs87}
\begin{equation}
I(q) = I(0)(1-q^{2}R_\mathrm{g}^{2}/3+\cdots).
\label{eq:guinierregion}
\end{equation}

\begin{figure}
\includegraphics[width=.9\columnwidth]{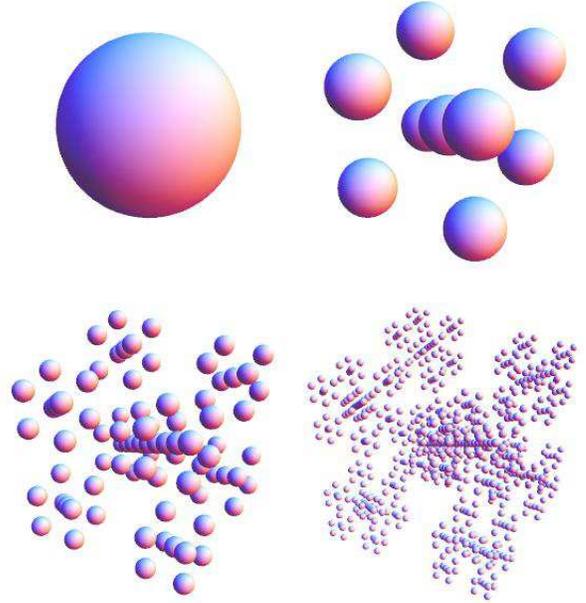}
\caption{\label{vid:gvf} (Color online) The initiator and first three iterations for the GSSVF at the scaling factor $\beta_\mathrm{s}=1/6$.  (See Ref.~\cite{[See Supplemental Material at ]video} for a video of the first iteration of GSSVF.)}
\end{figure}

\section{\label{sec:construction}{Construction of generalized self-similar Vicsek fractal}}

The construction of the generalized self-similar Vicsek fractal (GSSVF), embedded into 
three-dimensional space, is very similar to that of generalized Cantor fractals 
considered in Ref.~\cite{cherny210}.

We start with a cube with the edge $l_{0}$ and place in its center a ball of radius 
$l_{0}/2$. This is zero fractal iteration, called the \emph{initiator}. Let us choose the 
Cartesian coordinates: the origin lies in the cube center, and the axes are parallel to 
the cube edges. The iteration rule (\emph{generator}) is to replace the initial ball by 
nine smaller balls of radius $\beta_\mathrm{s} l_0/2$. The position of one ball is at 
the origin, while the centers of the eight other balls are shifted from the origin by 
the vectors
\begin{equation}
\bm{a}_{j}=\left\{ \pm \beta_\mathrm{t}l_{0}, \pm \beta_\mathrm{t}l_{0}, \pm
\beta_\mathrm{t}l_{0}\right\} \label{eq:shifts}
\end{equation}
with all the combinations of the signs, where
\begin{equation}
\beta_\mathrm{t} \equiv(1-\beta_\mathrm{s})/2.
\label{betat}
\end{equation}
The dimensionless positive parameter $\beta_\mathrm{s}$, called \emph{scaling factor}, 
obeys the condition $\beta_\mathrm{s} < \sqrt{3}/(\sqrt{3}+2)$ in order to avoid 
overlapping of the balls. Next iterations are obtained by performing an analogous 
operation to to each ball of the first iteration, and so on (see Fig.~\ref{vid:gvf}). 
Infinite number of iterations yields the ideal GSSVF.

Thus, the total number of balls at the $m$th iteration equals
\begin{equation}
N_{m}=9^{m},
\label{nm}
\end{equation}
and the corresponding radius is given by
\begin{equation}
r_{m}=\beta_\mathrm{s}^{m}l_{0}/2.
\label{radii}
\end{equation}
In accordance with the definition (\ref{eq:fd}), the fractal dimension can be calculated as (see also Appendix \ref{sec:hd})
\begin{equation}
D=\lim_{m\rightarrow \infty}{\frac{\log N_{m}}{\log (l_{0}/r_{m})}}=-\frac{2 \log 3}{\log \beta_\mathrm{s}}.
\label{massfractaldimension}
\end{equation}
Because of the restrictions imposed on the scaling factor, the fractal dimension of
GSSVF can vary from $0$ to $2.862\ldots$.

The above procedure is an obvious generalization of the original Vicsek fractal 
\cite{vicsek83}, which is constructed from the cubes by means of the scaling factor 
$\beta_\mathrm{s}=1/3$ and whose fractal dimension is equal to 2.

\section{\label{sec:formfactor}{The fractal form factor and structure factor}}
\begin{figure}[tbhp]
\centerline{\includegraphics[width=.9\columnwidth]{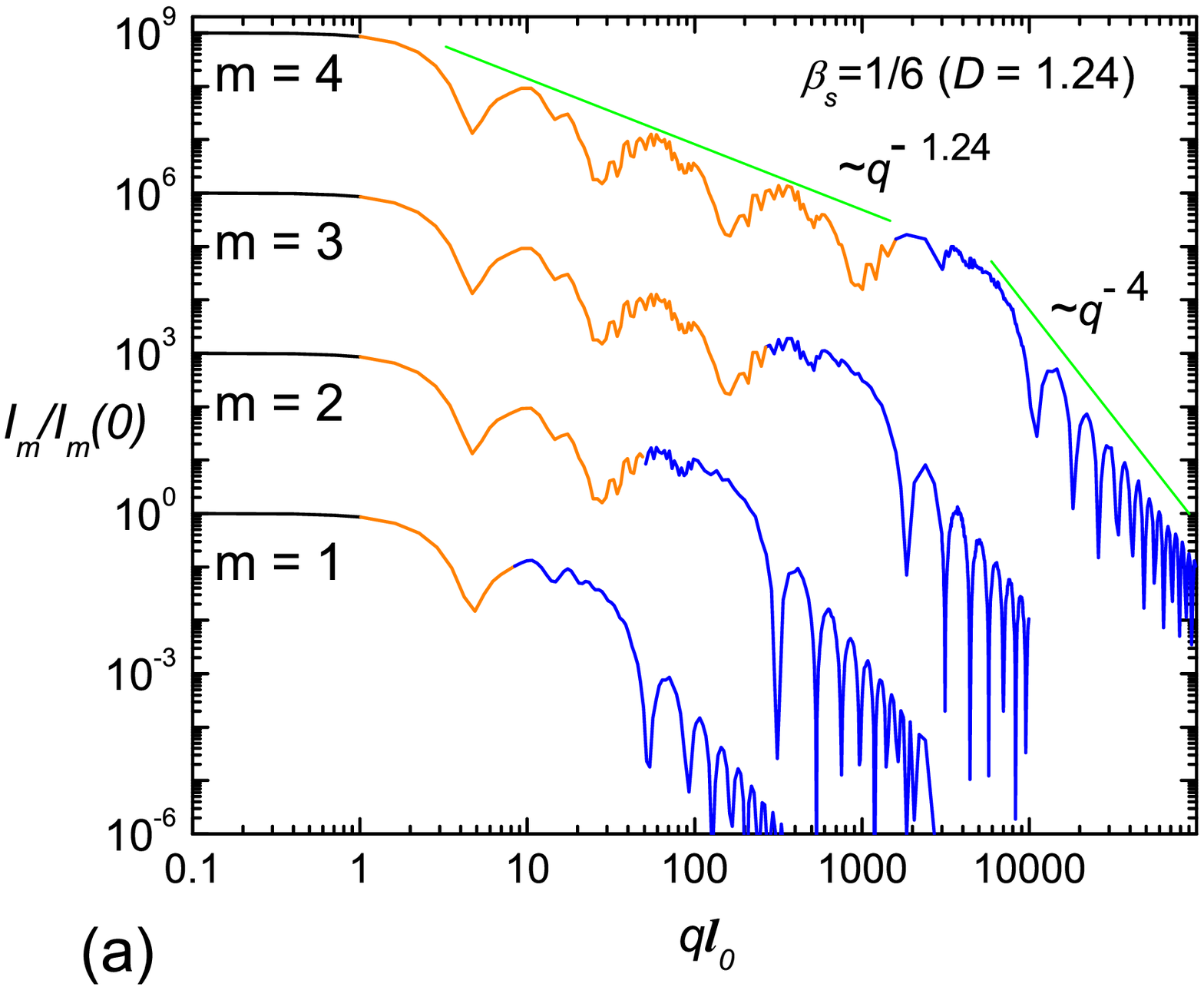}}
\centerline{\includegraphics[width=.925\columnwidth]{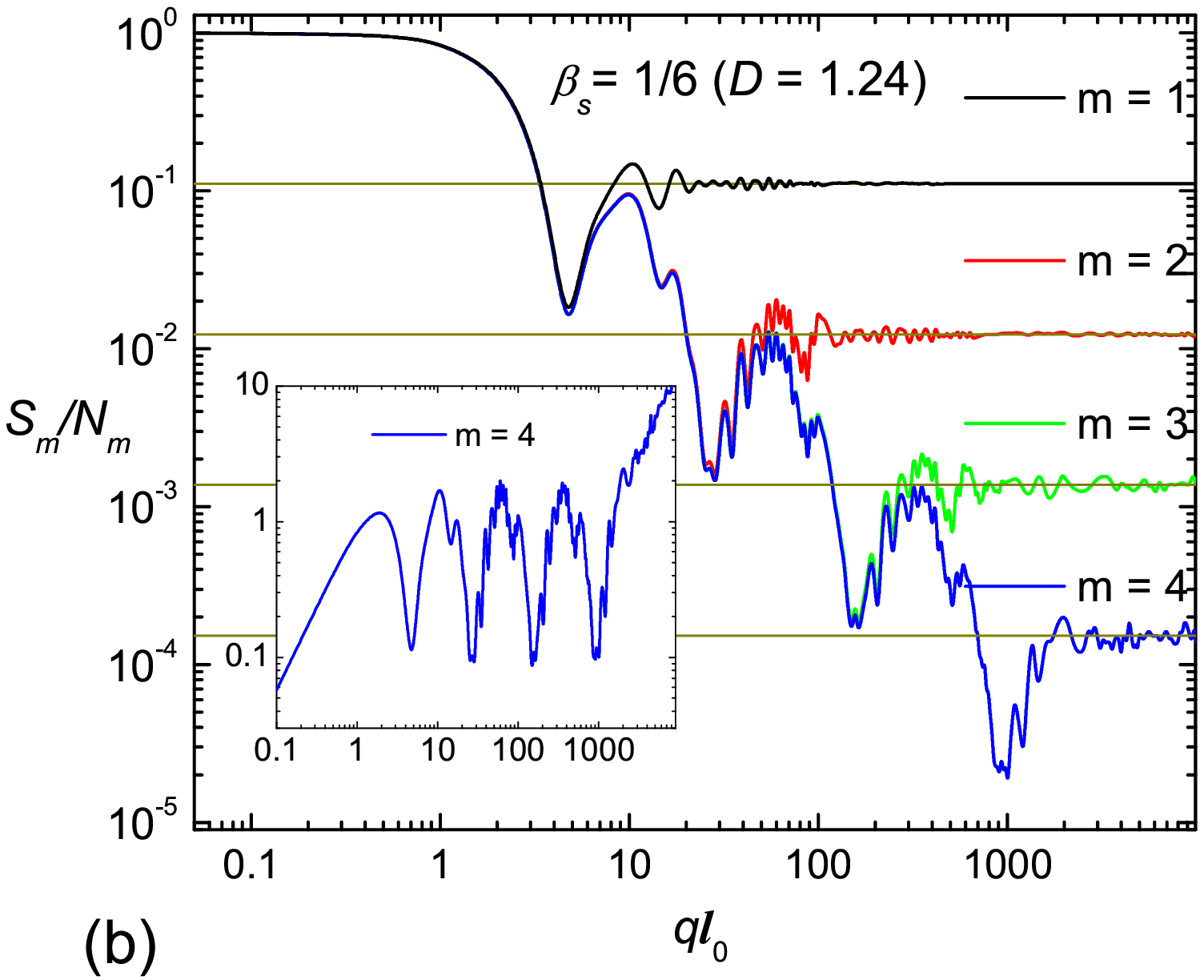}}
\centerline{\includegraphics[width=.925\columnwidth]{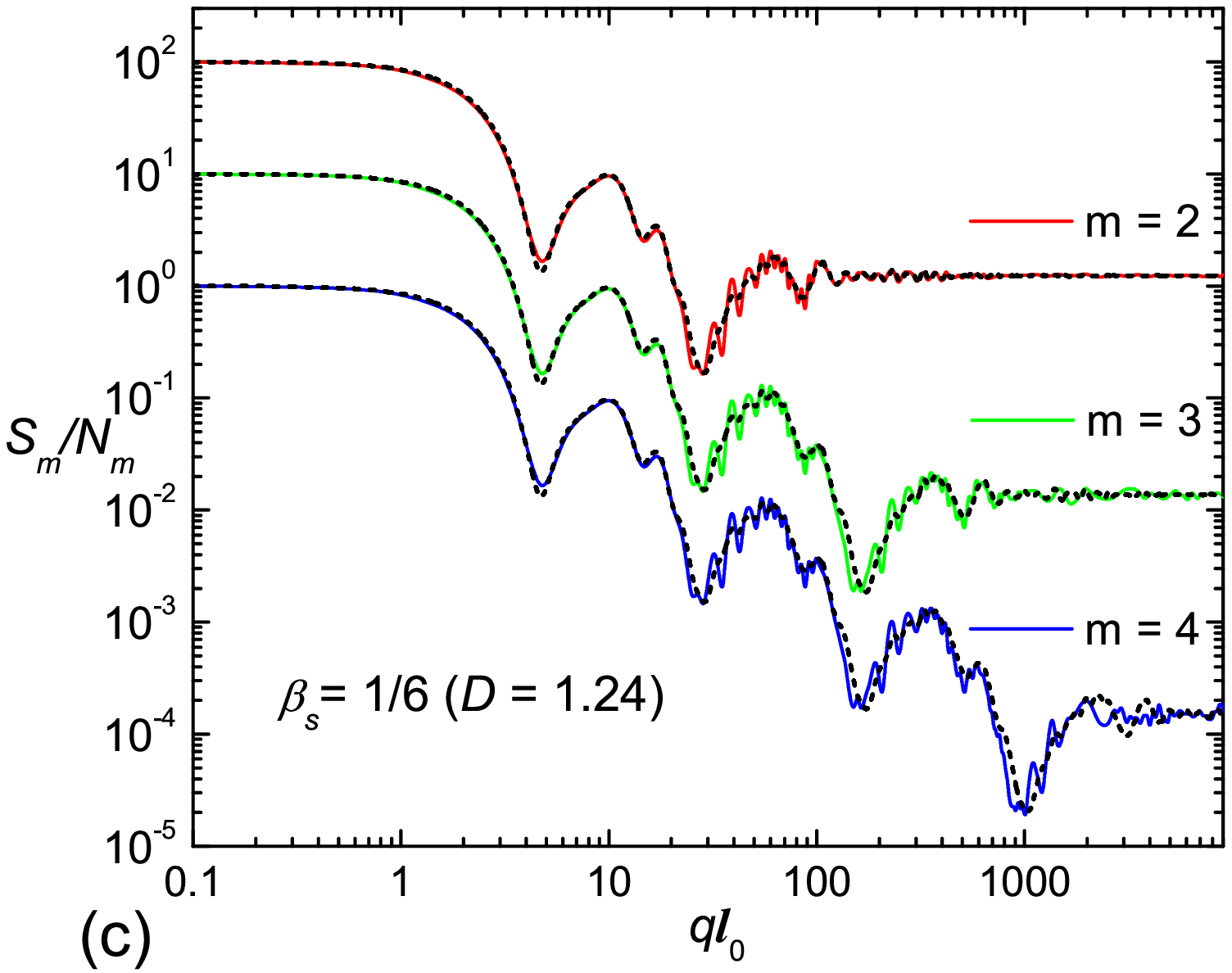}}
\caption{\label{fig:formfactors} (Color online) (a) Scattering intensity (\ref{eq:finalintensity}) for the first four iterations of monodisperse GSSVF. The scattering curve for the $m$th iteration is scaled up for clarity by the factor $10^{3(m-1)}$. Black at low q shows the Guinier regions; light gray (orange) represents fractal regions; dark gray (blue) at high $q$ Porod regions. (b) The fractal structure factor (\ref{eq:calcsf}) for the first four iterations. Inset: the quantity $S(q)(ql_0)^D/N_m$ versus $q l_0$, log-periodic in the fractal region. (c) The structure factor  (\ref{eq:calcsf}) (full lines) and its approximation (\ref{appstr}) (dashed lines) for three iterations. The values of structure factor for $m=3$ and $m=2$ are scaled-up by $10$ and $100$, respectively.}
\end{figure}
\subsection{\label{sec:analytexpr}{The analytical expressions}}

The standard method for calculating the fractal formfactor is the Debye formula 
\cite{debye15,gf55,fs87}, including the double integration over the scattering length 
density. Direct application of the formula could be extremely difficult, since the 
number of balls in the fractal increases exponentially with the iteration number. Here 
we apply an analytical method of calculating the fractal formfactor suggested in 
Refs.~\cite{kjems85:book,allain85:book,allain86} (see also Ref.~\cite{zygmund68:book}, 
where the Fourier transform of the Cantor measure was calculated). This scheme was 
developed afterwards in Refs.~\cite{sx86,schmidt91,lidar96,cherny210}. We give only the 
results here and refer the reader to our previous paper \cite{cherny210} for details 
of the derivation.

Once the number of balls at the $m$th iteration of GSSVF (\ref{nm}) and their radius (\ref{radii}) are known, the total fractal volume is given by
\begin{equation}
V_{m}=N_m\beta_{s}^{3m}V_0, \label{volume}
\end{equation}
where $V_0=\frac{4\pi}{3}(\frac{l_0}{2})^3$ is the volume of initial ball. Then the scattering intensity in zero angle $I_m(0)$ can be calculated with the help of Eq.~(\ref{intensityinzerogeneral}).

The formfactor of the $m$th generation is calculated analytically by means of the generative function, which is determined by the positions of the centers of balls inside the fractal for each iteration. For the GSSVF, the generative function reads
\begin{equation}
G_m(\bm{q})=\big[1+8\cos(u_{m}q_{x})\cos(u_{m}q_{y})\cos(u_{m}q_{z})\big]/9,
\label{eq:generativefunction}
\end{equation}
with $u_{m}\equiv l_{0}\beta_\mathrm{t}\beta_\mathrm{s}^{m-1}$ and $m=1,2,\cdots$. Here $\beta_\mathrm{t}$ is defined by Eq.~(\ref{betat}). If we put $G_{0}(\bm{q})\equiv 1$, one can express the form factor as
\begin{equation}
F_{m}(\bm{q})=F_{0}(r_m q)\prod_{i=0}^{m}G_{i}(\bm{q}), \label{eq:finalformfactor}
\end{equation}
for $m=0,1,2,\cdots$. The form factor of ball and its radius for the $m$th iteration are given by Eqs.~(\ref{ballformfactor}) and (\ref{radii}), respectively. Then SAS intensity is obtained with the help of Eq.~(\ref{intensitygeneral}) by averaging over the solid angle (\ref{aver})
\begin{equation}
I_{m}(q)/I_{m}(0)= \left\langle \left|F_{m}(\bm{q})\right|^{2}\right\rangle.
\label{eq:finalintensity}
\end{equation}
Equation~(\ref{eq:finalintensity}) gives the normalized scattering intensity as a function of $q l_0$ and the scaling factor $\beta_\mathrm{s}$.

Comparing Eqs.~(\ref{fqsq}) and (\ref{eq:finalformfactor}), one can conclude that the Fourier component of the density of ball centers in the $m$th iteration is given by
\begin{equation}
\rho_{\bm{q}}^{(m)}=N_m \prod_{i=0}^{m}G_{i}(\bm{q}).
\label{rhoqm}
\end{equation}
By substituting this expression into Eq.~(\ref{sq}), we obtain
\begin{equation}
S_{m}(q)/N_{m}=\Big\langle \prod_{i=1}^{m}|G_{i}(\bm{q})|^{2}\Big\rangle.
\label{eq:calcsf}
\end{equation}
It follows from Eqs.~(\ref{eq:finalformfactor}), (\ref{eq:finalintensity}), and (\ref{eq:calcsf}) that
\begin{equation}
I_{m}(q)/I_m(0) = |F_{0}(\beta_\mathrm{s}^{m}q l_0/2)|^{2} S_{m}(q)/N_{m},
\label{iqsqfq}
\end{equation}
which is in agreement with the general relation (\ref{intsq}).

\subsection{\label{sec:analysis}{Analysis and interpretation of the obtained results}}

First, let us discuss the physical meaning of the scattering amplitude (\ref{eq:finalformfactor}). It can be written in different forms
\begin{align}
F_{m}(\bm{q})&=G_{1}(\bm{q})F_{m-1}(\beta_\mathrm{s}\bm{q})\nonumber \\
&=G_{1}(\bm{q})G_{1}(\beta_\mathrm{s}\bm{q})F_{m-2}(\beta_\mathrm{s}^2\bm{q})=\cdots.
\label{ampl1}
\end{align}
The resulting amplitude can be understood as a sum of amplitudes of different clusters 
in the fractal, see Fig. \ref{vid:gvf}. As discussed in Sec.~\ref{sec:theory}, the 
choice of clusters is quite arbitrary. Thus, the amplitude (\ref{ampl1}) is a sum of 
$N_1=9$ amplitudes of $(m-1)$th fractal iterations, with $\rho_{\bm{q}}^{(1)}$ being the 
density Fourier component of their centers-of-masses. Or it is a sum of $N_2=9^2$  
scattering amplitudes of $(m-2)$th iterations, whose spatial positions are described by 
$\rho_{\bm{q}}^{(2)}$, and so on. Then we can represent the resulting intensity for each 
choice of clusters as a product of associated structure factor and formfactor $\langle 
|F_{m}(\bm{q})|^2\rangle \simeq S_1(q)\langle |F_{m-1}(\beta_\mathrm{s} 
\bm{q})|^2\rangle/N_1 \simeq S_1(q) S_1(\beta_\mathrm{s}q) 
\langle|F_{m-1}(\beta_\mathrm{s}^2\bm{q})|^2\rangle/N_1^2 \simeq \cdots \simeq S_1(q) 
S_1(\beta_\mathrm{s}q)$ $\cdots S_1(\beta_\mathrm{s}^{m-1}q) 
|F_{0}(\beta_\mathrm{s}^{m}q l_0/2)|^{2}/N_{m}$. Therefore, we arrive at the 
approximation for the fractal structure factor
\begin{equation}
S_m(q)\simeq S_1(q) S_1(\beta_\mathrm{s}q)\cdots S_1(\beta_\mathrm{s}^{m-1}q).
\label{appstr}
\end{equation}

The numerical results for the scattering intensities and related structure factor are shown in Fig.~\ref{fig:formfactors}. One can see a complex pattern of maxima and minima superimposed on a power-law decay (\ref{intensityreduced}), that is, the generalized power-law behavior of intensity, discussed in detail below. In the curves, one can separate out three different regions: the Guinier, fractal, and Porod regions.

\subsubsection{\label{Guinier} Guinier region}
The \emph{Guinier region}, as discussed in Sec.~\ref{sec:theory}, is
\begin{equation}
q\lesssim 1/l_0.
\label{guiner}
\end{equation}
In this region, the intensity is well described by the expansion (\ref{eq:guinierregion}). The intensity at zero angle is given by Eq.~(\ref{intensityinzerogeneral}) at the fractal volume (\ref{volume}), and $S_{m}(0) = N_{m}$. Expanding  the formfactor (\ref{eq:finalformfactor}) in power series in $q l_0$ and substituting the result into Eq.~(\ref{eq:finalintensity}) yield the fractal radius of gyration for the $m$th generation
\begin{equation}
R_\mathrm{g}=\sqrt{\beta_\mathrm{s}^{2m}R_{\mathrm{g}0}^2
+\frac{8}{3}\beta_\mathrm{t}^2\frac{1-\beta_\mathrm{s}^{2m}}{1-\beta_\mathrm{s}^2}l_0^2}.
\label{eq:rg}
\end{equation}
Here, for a uniform ball of radius $l_0/2$, the radius of gyration is given by $R_{\mathrm{g}0}=(3/5)^{1/2}l_0/2$. In the limit of high number of iterations (the ideal fractal), when $\beta_\mathrm{s}^{2m}\ll 1$, we obtain
\begin{equation}
R_\mathrm{g}=\frac{(8/3)^{1/2}\beta_\mathrm{t}l_{0}}{(1-\beta_\mathrm{s}^{2})^{1/2}}.
\label{eq:rgfinal}
\end{equation}

\subsubsection{\label{fractal} Fractal region}
The \emph{fractal region} in the momentum space is determined by the maximal and minimal distances between the ball centers, see Eq.~(\ref{fractalrangegeneral}). As can be seen from the fractal construction (Sec.~\ref{sec:construction}), these distances are of order $l_0$ and $l_{0}\beta_\mathrm{t}\beta^{m-1}_\mathrm{s}$, respectively. This yields for the fractal region
\begin{equation}
1/l_0\lesssim q \lesssim 1/(l_{0}\beta_\mathrm{t}\beta^{m-1}_\mathrm{s}).
 \label{frreg}
\end{equation}
The explicit analytical expressions 
(\ref{eq:generativefunction})-(\ref{eq:finalintensity}) allows us to check up these 
general estimations. For a given momentum, if the cosine argument in $G_{m+1}$ is much 
smaller than 1, then $G_{m+1}\simeq 1$, and further increasing of $m$ does not lead to 
an essential correction. Hence, the $m$th iteration reproduces the intensity that the 
ideal fractal would give at this momentum. As a result, the scattering amplitude of the 
$m$th iteration coincides with the amplitude of the ideal fractal within the region $q 
l_{0} \beta_\mathrm{t} \beta^{m-1}_\mathrm{s} \lesssim 1$, which is consistent with 
Eq.~(\ref{frreg}). Such a behaviour can be seen in Fig.~\ref{fig:formfactors} as the 
coincidence of various iterations in the fractal regions. In the fractal region, the 
value of normalized intensity is very close to that of structure factor
\begin{equation}
I_m(q)/I_m(0)\simeq S_m(q)/N_m,
\label{iqsq}
\end{equation}
since $F_0\simeq 1$ in Eq.~(\ref{iqsqfq}).

One can interpret the structure factor $S_1(q)$ by analogy with optics. The quantity 
$N_1 G_1(\bm{q})$ is nothing else but the amplitude, produced by nine interfering 
point-like particles of unit amplitude. The resulting scattering pattern 
$N_1^2\langle|G_1(\bm{q})|^2\rangle=S_1(q)N_1$ is the intensity, averaged over all 
directions of vector $\bm{q}$, see the representation (\ref{sqsum}). In optics, this 
average corresponds to diffraction with \emph{an entirely uncollimated beam}, which 
leads to the strong spatial incoherence. As a result, only the first minimum and maximum 
are quite distinguishable.  They are associated with out-of-phase ($\varphi=\pi$) and 
in-phase ($\varphi=2\pi$) interferences, respectively. The next minima ($\varphi=3\pi, 
5\pi,\cdots$) and maxima ($\varphi=4\pi, 6\pi,\cdots$) are not practically seen; 
however, they are responsible for the subsequent series of small oscillations. Besides, 
at $q=0$ we have a completely coherent diffraction (the intensity is equal to the 
\emph{squared number} of the points), and at large $q$, completely incoherent regime of 
geometrical optics (the intensity is equal to the \emph{number} of the points). By 
Eq.~(\ref{sqsum}), the phase is governed by distances between the points, which are of 
order $\sqrt{3} \beta_\mathrm{t}l_{0}$, and $\varphi\simeq \sqrt{3} \beta_\mathrm{t} q 
l_{0}$. Thus, we can summarize for the scattering from the first fractal iteration: 
$\varphi\lesssim 1$ implies entirely coherent regime, $\varphi\simeq \pi$ and 
$\varphi\simeq 2\pi$ correspond to minimum and maximum, respectively, and finally, the 
condition $\varphi\gtrsim 2\pi$ leads to entirely incoherent regime.

The behavior of the structure factor for arbitrary $m$th iteration can be described in the same manner. Each of the most pronounced minima or maxima corresponds to the interference of cluster amplitudes, where each cluster is a fractal iteration of the order $k=1,\cdots,m$. The most common distances between the center of masses of the clusters are equal to (see Sec.~\ref{sec:construction})
\begin{equation}
b_k=\sqrt{3}l_{0}\beta_\mathrm{t}\beta^{k-1}_\mathrm{s}.
\label{distclust}
\end{equation}
Then by analogy with the above considerations, the minima and maxima positions can be estimated from the condition that the distances (\ref{distclust}) equal $\pi/q$ and $2\pi/q$, respectively. Therefore, we obtain the conditions of minima
\begin{align}
q_{k}l_{0} \approx \frac{\pi}{\sqrt{3}\beta_\mathrm{t}\beta_\mathrm{s}^{k-1}},\quad k=1,\cdots,m,
\label{eq:minimapositions}
\end{align}
and the conditions of maxima
\begin{align}
q_{k}l_{0} \approx \frac{2\pi}{\sqrt{3}\beta_\mathrm{t}\beta_\mathrm{s}^{k-1}},\quad k=1,\cdots,m.
\label{eq:maximapositions}
\end{align}
These relations are satisfied with a good accuracy, see Fig.~\ref{fig:formfactors}. 
Thus, the number of minima is equal to the number of fractal iteration. The same is 
true for the number of maxima.

To explain the generalized power-law behavior, one can use the approximate expression (\ref{appstr}), which works quite well (Fig.~\ref{fig:formfactors}c). Note that for a given range $1/(\sqrt{3} \beta_\mathrm{t} \beta_\mathrm{s}^{k-1}l_0)\lesssim q \lesssim 1/(\sqrt{3} \beta_\mathrm{t} \beta_\mathrm{s}^{k}l_0)$, only one term $S_1(\beta_\mathrm{s}^{k-1}q)$ in the product (\ref{appstr}) has a non-trivial behavior. The other terms are nearly equal $N_1=9$ (the coherent regime for the terms on the right) or approximately equal one (the incoherent regime for the terms on the left). This physically means that the diffraction pattern in this range is produced by the interference of only one group of subunits, which are the $k$th fractal iterations. Hence, by increasing the argument $1/\beta_\mathrm{s}$ times, one term on the right in the product is forced from the coherent regime into the incoherent one, thus reducing the total value of the structure factor $N_1=9$ times. Therefore, we arrive at the equation
\begin{equation}
S_m(q/\beta_\mathrm{s})\simeq \beta_\mathrm{s}^D S_{m}(q),
\label{sqaffin}
\end{equation}
valid for $1/(\sqrt{3} \beta_\mathrm{t}l_0)\lesssim q \lesssim 1/(\sqrt{3} \beta_\mathrm{t} \beta_\mathrm{s}^{m-2}l_0)$. Here we use the equation $N_1=1/\beta_\mathrm{s}^D$, which follows from the definition of fractal dimension (\ref{massfractaldimension}). Equation (\ref{sqaffin}) in conjunction with the relation (\ref{iqsq}) yields
\begin{equation}
I_m(q/\beta_\mathrm{s})(q/\beta_\mathrm{s})^D\simeq I_{m}(q)q^D
\label{iqlog}
\end{equation}
for the same region. This means that within the region, the functions $S_{m}(q)q^D$ and 
$I_{m}(q)q^D$ are approximately \emph{log-periodic} with the period, equal to the 
inverse scaling factor $1/\beta_\mathrm{s}$ [see Fig.~\ref{fig:formfactors}b].

\subsubsection{\label{Porod} Porod region}

\emph{Beyond} the fractal region
\begin{equation}
q\gtrsim 1/(\beta_\mathrm{t}\beta_\mathrm{s}^{m-1}l_0),
\label{beyondfr}
\end{equation}
$S_{m}(q)\simeq 1$, as discussed above, and we have in accordance with Eq.~(\ref{iqsqfq})
\begin{equation}
I_{m}(q)/I_m(0) = |F_{0}(\beta_\mathrm{s}^{m}q l_0/2)|^{2}/N_{m}.
\label{iqbeyond}
\end{equation}
It follows that beyond the fractal region, the scattering intensity resembles the intensity of the initiator, \emph{i.e.}, a ball in the present case. In particular, the maxima of the curve obey Porod's law $q^{-4}$ \cite{gf55,fs87} in the \emph{Porod region}
\begin{equation}
q\gtrsim 1/(\beta_\mathrm{s}^{m}l_0), \label{porodrg}
\end{equation}
whose lower border is of order the upper edge of fractal region. Such a behavior can be seen in Fig.~\ref{fig:formfactors}a. However, if the radius of initiator is much smaller than $l_0$ then we have an intermediate region, where $I_{m}(q)/I_m(0) \simeq 1/N_{m}$. So, a well or even ``shelf" appears in the normalized intensity between the fractal and Porod regions near the volume $1/N_m$ \cite{cherny210}.

The generalized power-law behavior of GSSVF is consistent with the SAS from other types 
of deterministic fractals reported in the literature \cite{sx86, schmidt91, cherny110, cherny210}. In the 
paper \cite{sx86}, the authors studied SAS from Menger sponges and another type of 
fractals and found some general scattering properties of deterministic fractals:  the 
generalized power-law behavior of the intensities and log-periodicity of the main maxima 
and minima. However, the method of calculations does not allow the authors to 
distinguish between the fractal structure factor and the total scattering intensity; for 
instance, the correct asymptotics of structure factor was not found, and the positions 
of the minima and maxima were not estimated. In the paper \cite{lidar96}, self-affine 
characteristics of the scattering curves, like Eq.~(\ref{sqaffin}), were investigated; 
however, the author studied the exact anisotropic formfactors like 
Eq.~(\ref{eq:finalformfactor}), which were \emph{not} averaged over all directions of 
momentum. We believe that the log-periodicity, even approximate, takes place only after 
averaging the scattering curves over the angles. We discuss this important issue in 
Sec.~\ref{sec:geometry} below.

\section{\label{sec:polydisperse}{Polydisperse formfactor and structure factor}}
\begin{figure}[tbhp]
\centerline{\includegraphics[width=.9\columnwidth]{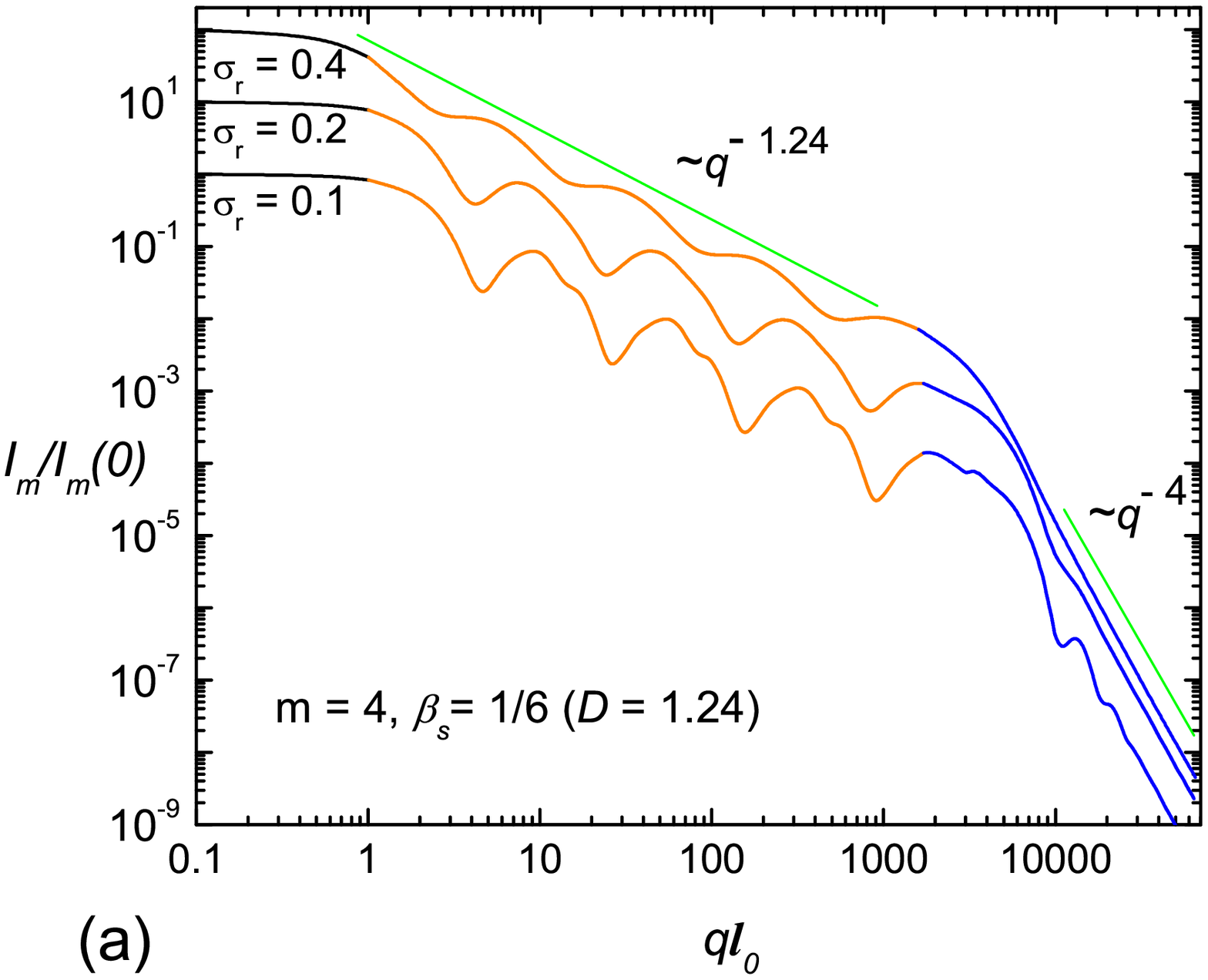}}
\centerline{\includegraphics[width=.92\columnwidth]{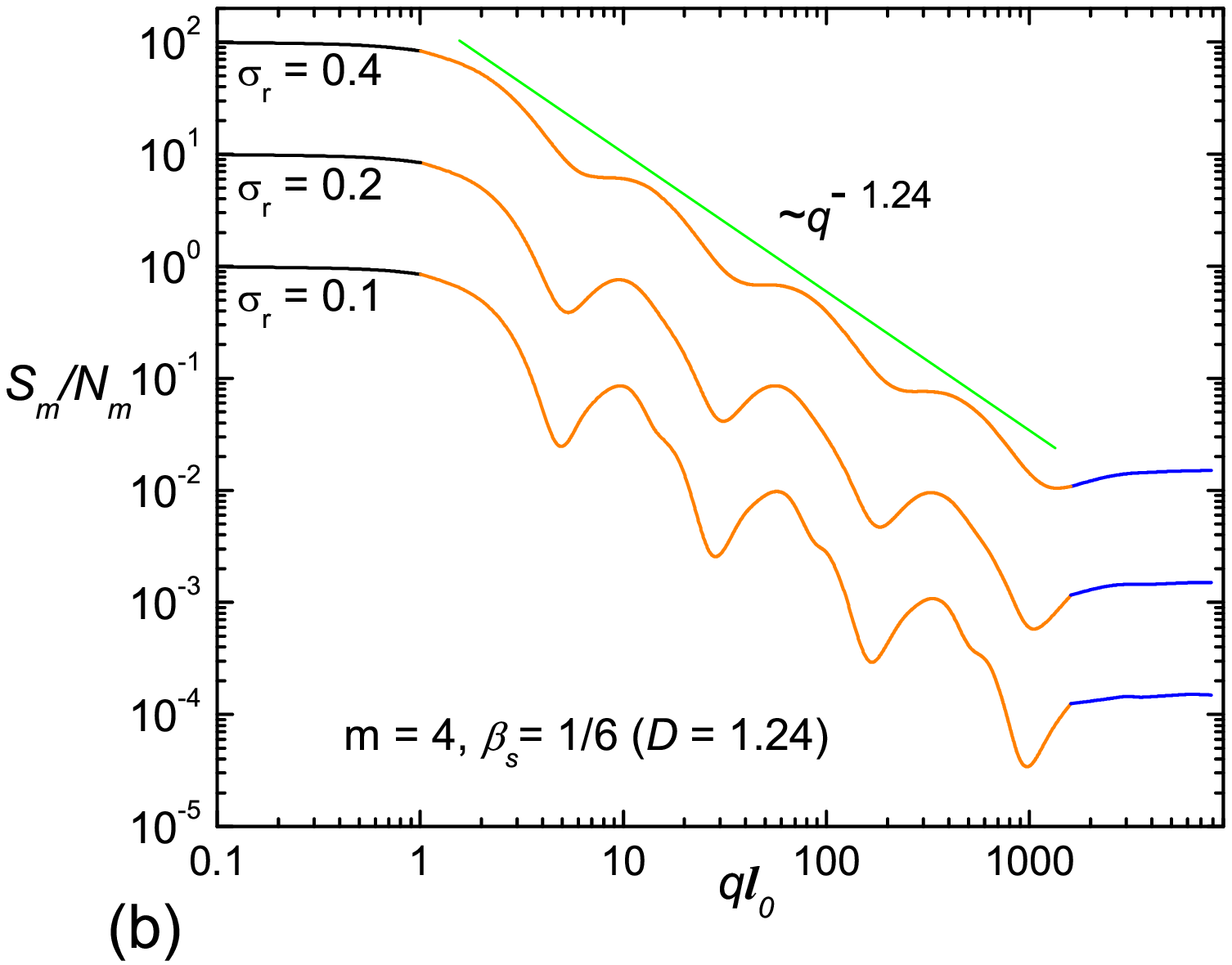}}
\centerline{\includegraphics[width=.9\columnwidth]{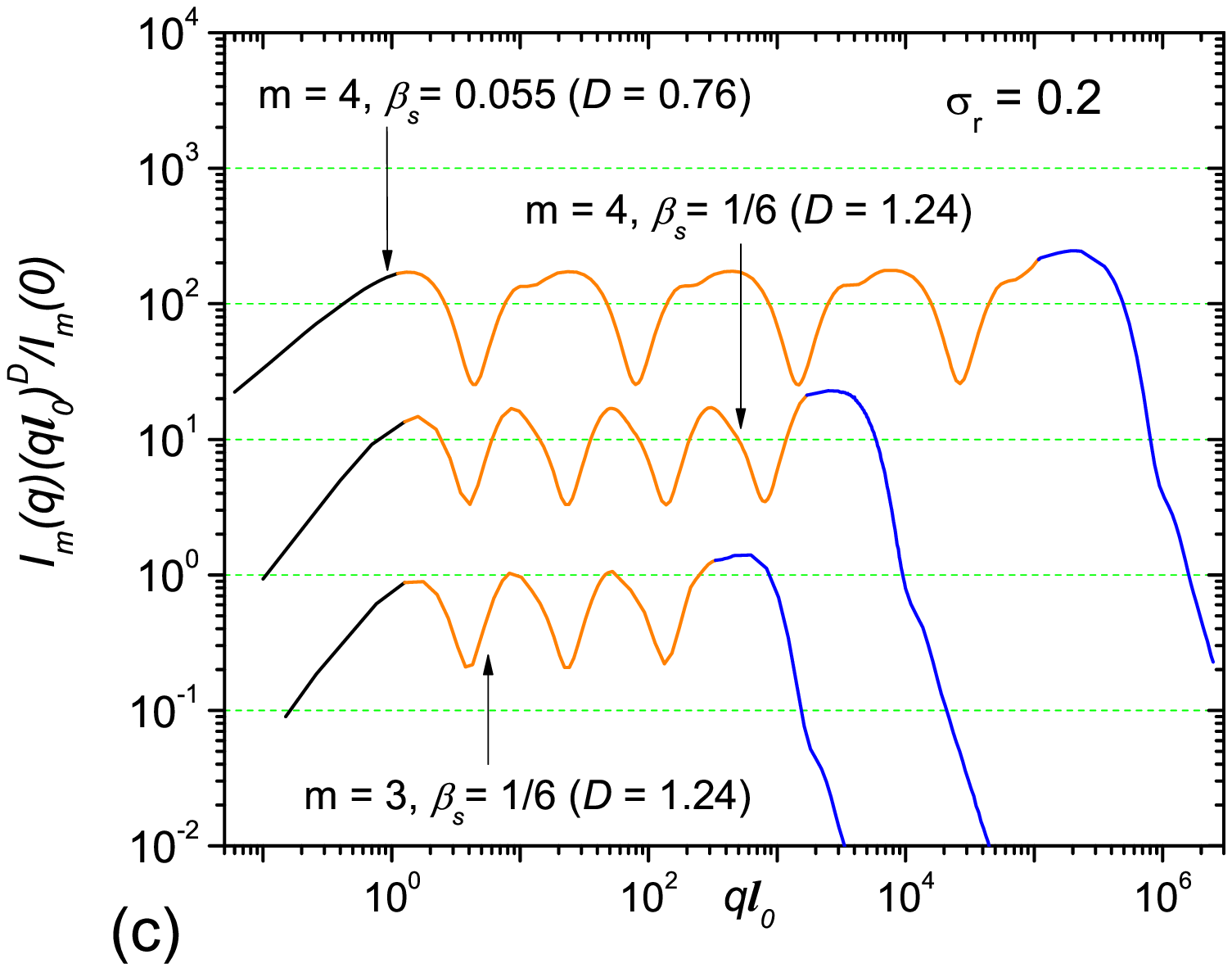}}
\caption{\label{fig:polydispersity} (color online) Influence of polydispersity on the scattering curves: (a) intensity (\ref{eq:ffpolydisperse}) and (b) structure factor (\ref{sqpoly}) for various values of the relative variance $\sigma_\mathrm{r}$, given by Eq.~(\ref{eq:lengthandvariance}). Scattering curves for $\sigma_\mathrm{r}=0.2$ and $0.4$ are scaled-up for clarity by $10$ and $100$, respectively. (c) The quantity $I(q)q^D$ clearly shows the log-periodicity in the fractal region. The period in the log-scale is equal to $\log_{10} (1/\beta_\mathrm{s})$, where $\beta_\mathrm{s}$ is the scaling factor of fractal. The colors are the same as in Fig.~\ref{fig:formfactors}.}
\end{figure}

In a real physical system, scatterers almost always have different sizes. Therefore, a more realistic description should involve \emph{polydispersity}. This means that we deal with a set of fractals with various sizes and forms, in general. Here we can consider a kind of polydispersity, an ensemble of GSSVF with different sizes $l$ (that is, $l$ is the length of initial cube, see Sec.~\ref{sec:construction}). Note that in the previous sections we denote the fractal size as $l_0$, while here and below $l_0$ is the \emph{mean} value of fractal sizes over the ensemble. The distribution function $D_\mathrm{N}(l)$ of the scatterer sizes is defined in such a way that $D_\mathrm{N}(l)dl$ gives the probability of finding a fractal whose size falls within the interval $(l,l+dl)$. We consider here one of the most common distribution function, the log-normal distribution, given by
\begin{equation}
D_\mathrm{N}(l)=\frac{1}{\sigma l
(2\pi)^{1/2}}\exp\bigg(-\frac{\big[\log(l/l_{0})+\sigma^{2}/2\big]^2}{2\sigma^{2}}\bigg),
\label{eq:distribution}
\end{equation}
where $\sigma=[\log(1+\sigma_\mathrm{r}^{2})]^{\frac{1}{2}}$. The quantities $l_0$ and $\sigma_\mathrm{r}$ are the mean length and its relative variance
\begin{equation}
l_{0}\equiv\left\langle l\right\rangle_{D},\quad \sigma_\mathrm{r}\equiv\big(\langle
l^{2}\rangle_{D}-l_{0}^{2}\big)^{1/2}/l_{0}, \label{eq:lengthandvariance}
\end{equation}
where $\left\langle \cdots \right\rangle_{D} \equiv \int_{0}^{\infty} \cdots D_\mathrm{N}(l)\d l$.

As in the previous sections, we assume that spatial positions of different fractals are uncorrelated. Hence, the resulted intensity is the average of the intensity (\ref{intensitygeneral}) over the distribution function (\ref{eq:distribution})
\begin{equation}
I_{m}(q)=n |\Delta\rho|^{2}\int_{0}^{\infty}\left\langle
|F_{m}(\bm{q})|^{2}\right\rangle V_{m}^2(l)D_\mathrm{N}(l)\mathrm{d}l,
\label{eq:ffpolydisperse}
\end{equation}
where the volume and amplitude of monodisperse fractal are given by Eqs.~(\ref{volume}) and (\ref{eq:finalformfactor}), respectively. The fractal structure factor in the presence of polydispersity can be calculated in the same manner but without the term $V_{m}^2(l)$
\begin{equation}
S_{m}(q)=N_{m}\int_{0}^{\infty}\Big\langle \prod_{i=1}^{m}|G_{i}(\bm{q})|^{2}\Big\rangle
D_\mathrm{N}(l)\mathrm{d}l,
\label{sqpoly}
\end{equation}
where $G_{i}$ is given by Eq.~(\ref{eq:generativefunction}). Equations (\ref{eq:ffpolydisperse}) and (\ref{sqpoly}) correspond to the different statistical ensembles. Indeed, the former averaging  assumes that the size of balls, constituting the fractal, equals $l/2$ and thus it is also distributed in accordance with Eq.~(\ref{eq:distribution}). The latter averaging assumes that all the distances between the ball centers inside the fractal are changed when the fractal size varies, but the radius of the balls is invariable. In this case, the structure factor becomes clearly defined even in the presence of polydispersity.

Figure~\ref{fig:polydispersity} shows the influence of polydispersity on the obtained scattering intensity (\ref{eq:ffpolydisperse}) and structure factor (\ref{sqpoly}).  As expected, the small oscillations, present in the monodisperse case (see Fig.~\ref{fig:formfactors}), are now smeared out, and the scattering curves became smoother. One can see that the smoothness increases with growing the width of the distribution function, which is controlled by the relative variance $\sigma_\mathrm{r}$. Nevertheless, the exponent of the generalized power-law dependence remains unchanged and equal $D$ in the fractal region (\ref{frreg}). In the Porod region (\ref{porodrg}), the value of exponent still equals $-4$. The structure factor is close to one beyond the fractal region (\ref{beyondfr}). The logarithmic periodicity of the quantity $I(q)q^D$ becomes even more pronounced in comparison with the monodispersive fractals, discussed in Sec.~\ref{fractal}.


\section{\label{sec:geometry}{Pair distribution function and pair distance distribution function}}
\begin{figure}[tbhp]
\centerline{\includegraphics[width=.825\columnwidth]{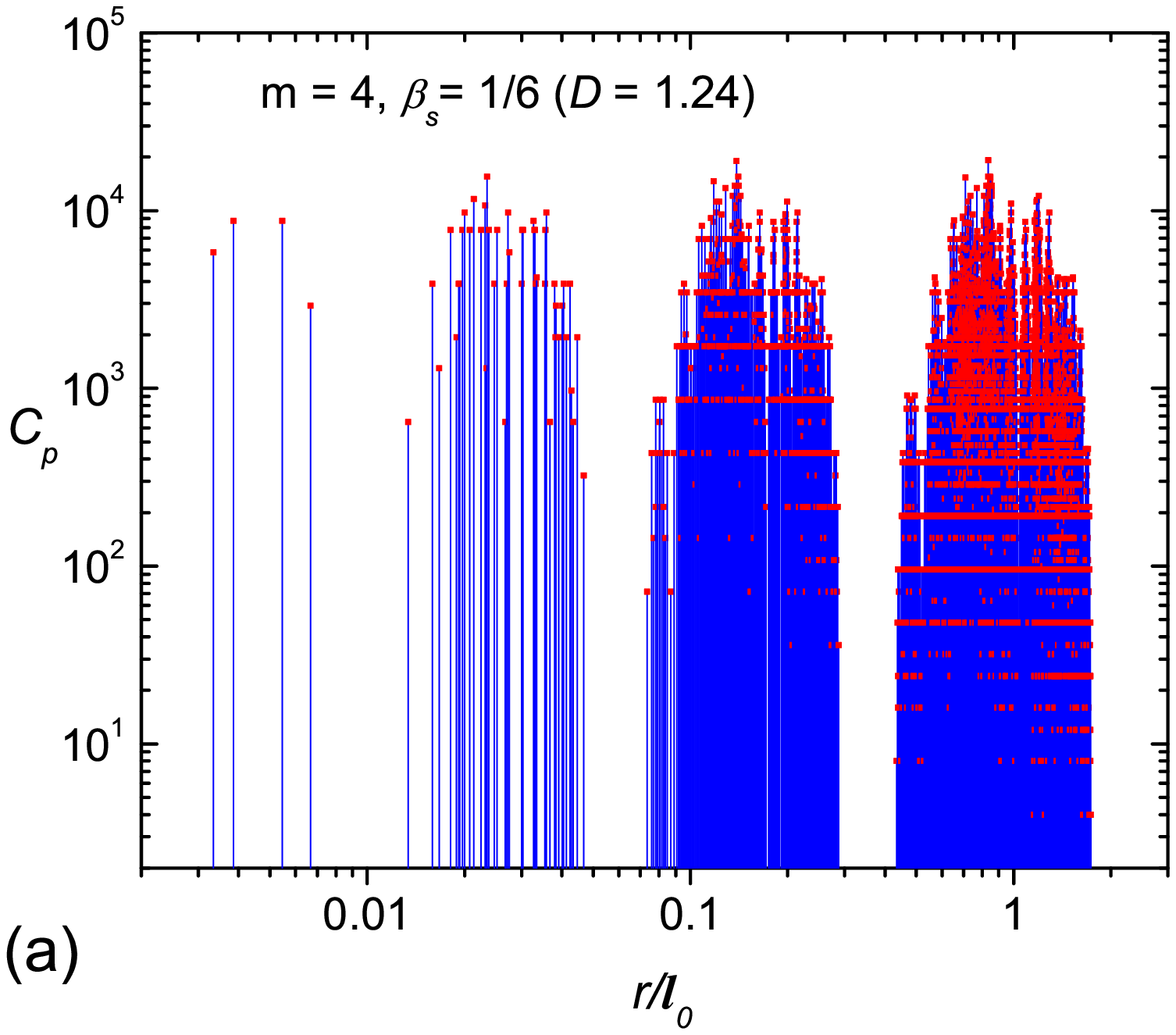}}
\centerline{\includegraphics[width=.85\columnwidth]{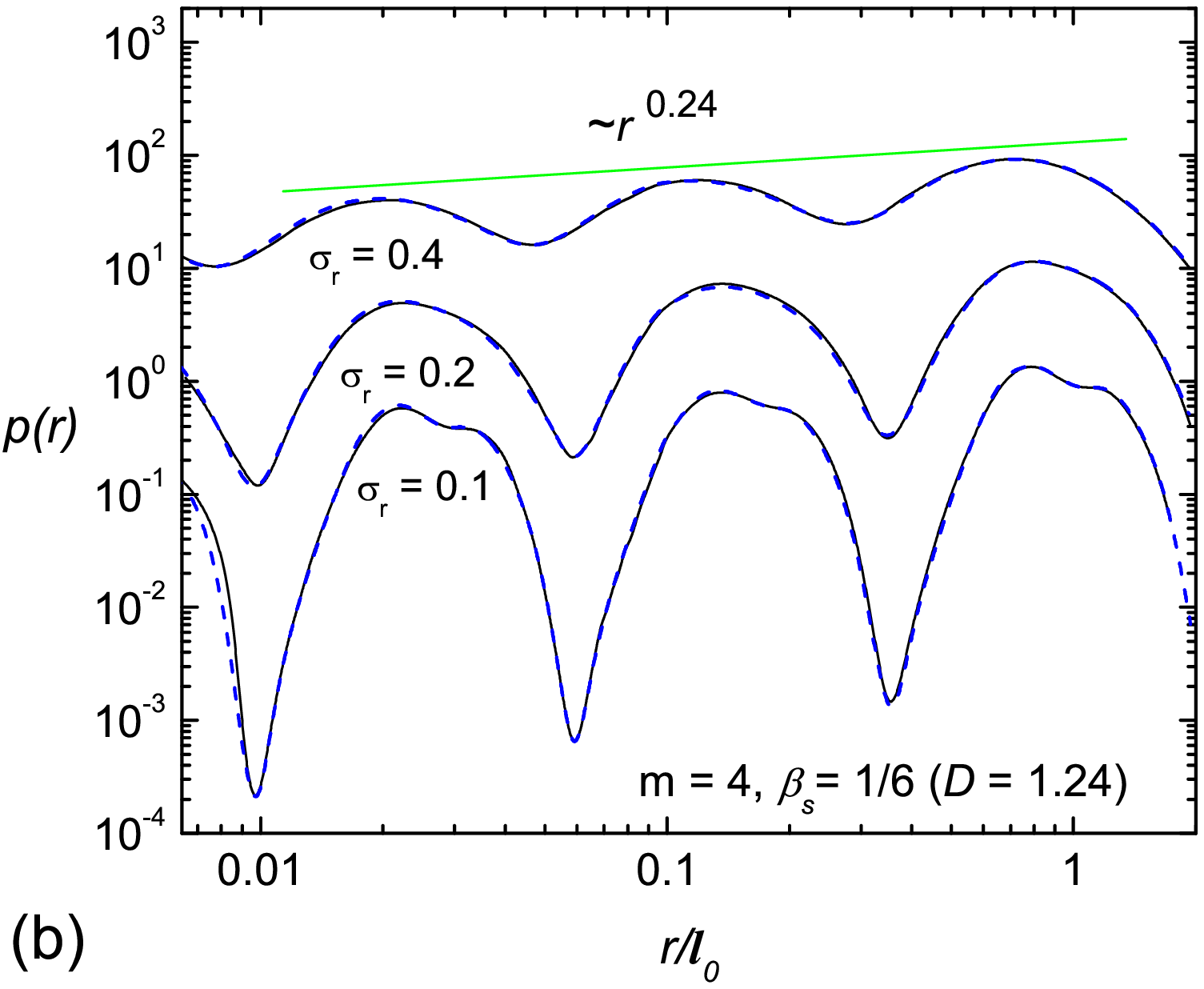}}
\centerline{\includegraphics[width=.875\columnwidth]{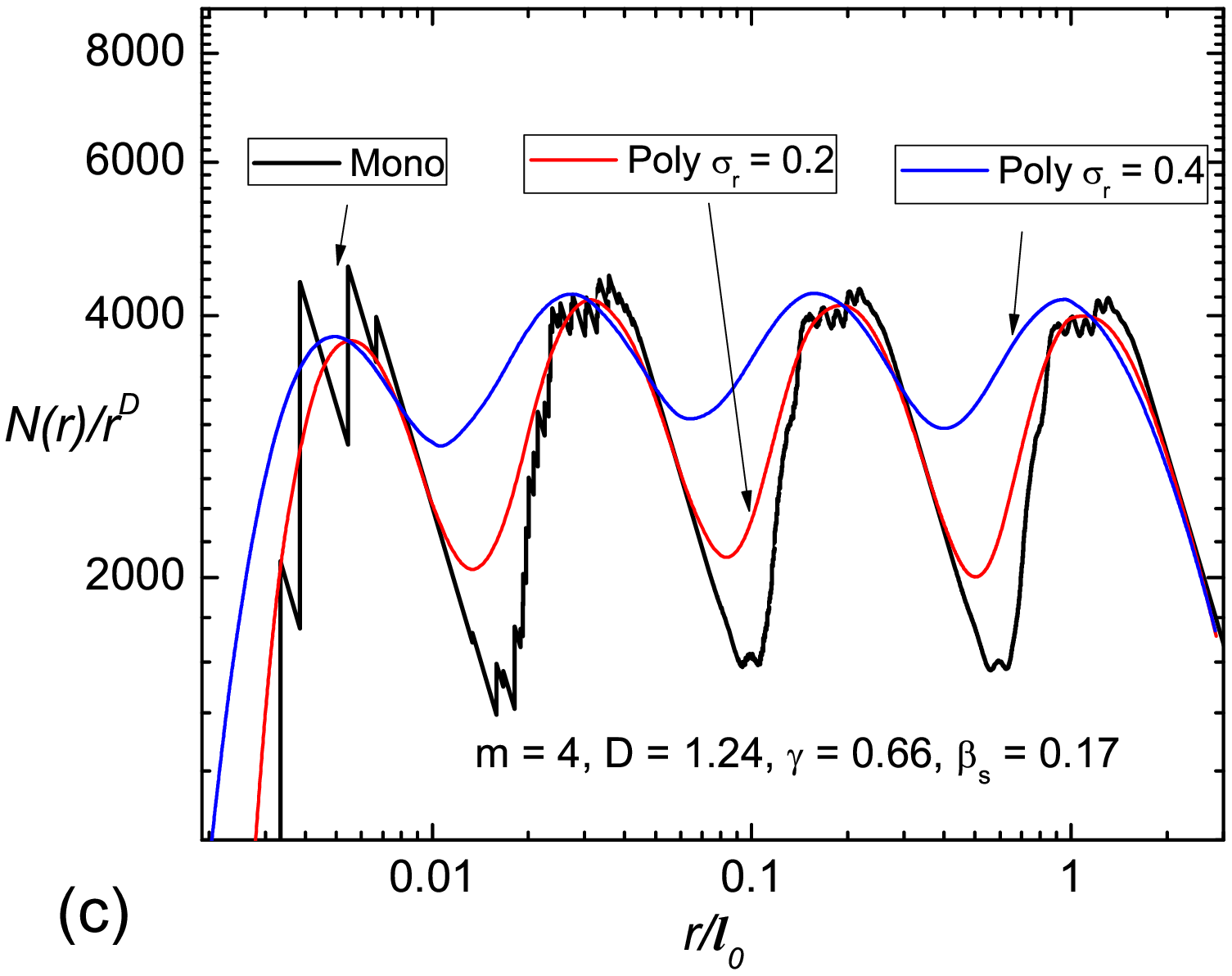}}
\caption{\label{fig:discreteprs} (Color online) (a) The coefficients $C_p$ in the expression (\ref{eq:defpr}) for the pair distribution function of GSSVF. The centers of groups are the distances between the clusters (\ref{distclust}), and the period on the log-scale is equal to $\log_{10} (1/\beta_\mathrm{s})$. (b) The influence of polydispersity on the pair distance distribution function. Scattering curves for the distribution width $\sigma_\mathrm{r}=0.2,0.4$ are scaled-up for clarity by $10$ and $10^2$ respectively. Solid (black) lines show the solutions obtained with Eq.~(\ref{eq:prsqav}) from the polydisperse structure factor of Fig.~\ref{fig:polydispersity}b, dashed (blue) lines represent its direct calculation (\ref{eq:prav}). The shift of maxima in comparison to Eq.~(\ref{distclust}) is due to a polydispersity effect. The exponent of the generalized power-law behavior for $p(r)$ is equal to $D-1$, see Eq.~(\ref{eq:pexp}). (c) The log-periodic function $N(r)/r^D$ with the same log-period. The quantity $N(r)$ (\ref{nrpr}) is proportional to the ``mass", or measure, enclosed in the imaginary sphere of radius $r$, which is centered on fractal.}
\end{figure}

\subsection{\label{sec:gendef}General definitions}

The formfactor and related structure factor are intimately connected to spatial correlations of subunits, composing the investigated system, and the main problem of SAS is to extract information about the real-space structure from the scattering data.

Let us consider first a monodisperse fractal, composed of balls of the same radius. The general expression (\ref{sqsum}) can be rewritten in the form
\begin{equation}
S(q)=1+\frac{2}{N}\sum_{1 \leqslant k < j \leqslant N}\frac{\sin qr_{jk}}{qr_{jk}},
\label{eq:structurefactor}
\end{equation}
where $r_{jk} \equiv \left|\bm{r}_j-\bm{r}_k\right|$ are the relative distances between 
the ball centers. Deriving this last relation, we use the formula 
$\langle\exp(i\bm{a}\cdot\bm{q})\rangle=\sin(aq)/(aq)$, which follows from 
equation~(\ref{aver}). In Eq.~(\ref{eq:structurefactor}), the sum contains many equals 
terms, because the distances separating different points can coincide. One can introduce 
the probability density of finding the distance $r$ between the centers of two arbitrarily 
taken balls \emph{inside} the fractal:
\begin{align}
p(r)&\equiv \frac{2}{N(N-1)}\sum_{k<j}\delta(r-r_{jk}) \nonumber\\
    &=\frac{2}{N(N-1)}\sum_{r_p} C_p\, \delta(r-r_{p}),
\label{eq:defpr}
\end{align}
where $C_{p}$ are the numbers of distances separated by $r_{p}$. The dependence on the fractal length $l$ appears through $r_{p}=l z_{p}$, where the distances $z_{p}$ correspond to the fractal of unit length. The quantity (\ref{eq:defpr}) is called \emph{pair distance distribution function}. In terms of this function, Eq.~(\ref{eq:structurefactor}) reads
\begin{equation}
S(q)=1+(N-1)\int_{0}^{+\infty}\d r\,p(r)\frac{\sin qr}{qr}. \label{srpr}
\end{equation}
By performing the inverse Fourier sine transform, we obtain
\begin{equation}
p(r)=\frac{2}{\pi}\int_0^{+\infty}\frac{S(q)-1}{N-1}qr\sin q r \,\d q. \label{eq:prsqav}
\end{equation}

Let us choose an arbitrary ball in the fractal and consider a spherical layer of radius $r$ and width $\d r$, whose center coincides with the center of the chosen ball. It follows from the definition (\ref{eq:defpr}) that $\d N=(N-1)p(r)\d r$ gives the average number of other ball centers, which lay within the spherical layer. Then the total average number of other balls in the sphere of radius $r$ is given by
\begin{equation}
N(r)=(N-1)\int_{0}^{r} p(r')\d r'.
\label{nrpr}
\end{equation}
Conversely, $(N-1) p(r)={\d N(r)}/{\d r}$. The quantity $N(r)$ (\ref{nrpr}) is proportional to the ``mass", or measure, enclosed in the imaginary sphere of radius $r$, which is centered on fractal, see the discussion in Appendix \ref{sec:hd}. The relation (\ref{nrpr}) gives a more precise definition of $N(r)$, discussed in Sec.~\ref{sec:intro}. By substituting Eq.~(\ref{eq:prsqav}) into Eq.~(\ref{nrpr}), we arrive at the relation between $N(r)$ and the fractal structure factor
\begin{equation}
N(r)=\frac{2}{\pi}\int_0^{+\infty}\frac{S(q)-1}{q}[\sin(qr) -qr \cos(qr)]\,\d q.
\label{nrsq}
\end{equation}

The polydispersity can be taken into account by averaging the above equations (\ref{eq:defpr})-(\ref{nrsq}) over the distribution (\ref{eq:distribution}). All the relations take the same form but with the replacements $S(q)\to \langle S(q)\rangle_D$, $N(r)\to \langle N(r)\rangle_D$, $p(r)\to \langle p(r)\rangle_D$. The last quantity can be calculated explicitly
\begin{equation}
\langle p(r)\rangle_{D}=\frac{2}{N(N-1)}\sum_{p}
\frac{C_{p}}{z_{p}}D_\mathrm{N}\left(\frac{r}{z_{p}}\right). \label{eq:prav}
\end{equation}
For a finite iteration, the total number of balls $N$ is given by Eq.~(\ref{nm}).

For describing spatial particle correlations, one can also use \emph{pair distribution function} \cite{defgr} $g(r)$, which is directly related to the pair distance distribution function (\ref{eq:defpr})
\begin{equation}
g(r)\equiv\frac{p(r)l^{3}}{4\pi r^2},
\label{eq:gandp}
\end{equation}
where $l^3$ is the total volume of fractal. As discussed in Sec.~\ref{sec:intro}, the pair distribution function has a transparent physical interpretation:  it is nothing else but the conditional probability density, because it gives the probability density to find a particle at the distance $r$ from another particle, provided a position of the latter particle is given. As follows from the discussion below Eq.~(\ref{eq:prsqav}), the number of particles within the spherical layer is indeed given by $\d N=n g(r)\d V$. Here $n=(N-1)/l^3$ is the average particle density in the fractal (if we neglect the difference between $N$ and $N-1$ for a large number of particles), and $\d V=4\pi r^2 \d r$ is the layer volume.

Note that Eq.~(\ref{eq:gandp}) is consistent with the other definition $g(r)\equiv 
l^3/[N(N-1)] \langle \sum_{i\not=j}\delta(\bm{r} - \bm{r}_{ij})\rangle$, more common in 
the literature (see, e.g., Ref.~\cite{march67:book}). Here the brackets denote the 
average over all the directions of $\bm{r}$, and $\bm{r}_{ij}$ are the relative 
positions of two particles. By using the average of the $\delta$-function 
$\langle\delta(\bm{r} - \bm{a})\rangle = \delta(r-a) /(4\pi r^2)$, we derive 
Eq.~(\ref{eq:gandp}).

\subsection{\label{sec:realres} Analysis of results}

The real-space characteristics are shown in Fig.~\ref{fig:discreteprs}. Figure \ref{fig:discreteprs}a represents the coefficients in the expression (\ref{eq:defpr}) for the pair distance distribution function. They are found numerically for the fourth iteration by a simple combinatoric analysis. The self-similarity of GSSVF manifests itself in the periodicity of the groups of distances on the logarithmic scale. As expected, the centers of groups are the distances between the clusters (\ref{distclust}), discussed in Sec.~\ref{fractal}, and the period in the log-scale is related to the scaling parameter through $\log_{10} (1/\beta_\mathrm{s})$.

When the polydispersity is present, the pair distance distribution function can be calculated from Eq.~(\ref{eq:prsqav}) with the polydisperse structure factor (\ref{sqpoly}) or directly from Eq.~(\ref{eq:prav}) with known quantities $C_p$ and $z_p$. We obtain a very good agreement between the two formulas, see Fig.~\ref{fig:discreteprs}b. Small differences appear at the low distances because of the finite upper integration limit in Eq.~(\ref{eq:prsqav}), which we have to use in numerical calculations. Equation (\ref{eq:prsqav}) is nevertheless preferable for large fractal iterations, since it surpasses the inconvenience of calculating the exponentially increasing number of distances. This gives a nearly independent computation time with increasing the iteration number.

The simple power law (\ref{grpower}) for $g(r)$ implies that $p(r)\varpropto r^{D-1}$ due to the relation (\ref{eq:gandp}). Instead, one can see the generalized power-law behavior in Fig.~\ref{fig:discreteprs}b. Moreover, the quantity $p(r)/r^{D-1}$ is log-periodic in the fractal region in the real space, which results from the self-affinity relation (\ref{sqaffin}) for the structure factor. Indeed, the main contribution in the integral of Eq.~(\ref{eq:prsqav}) comes from the fractal region in momentum space, where $S(q)\gg 1$ and Eq.~(\ref{sqaffin}) is satisfied. Then, equating $S(q)-1$ to $S(q)$ in the integral of Eq.~(\ref{eq:prsqav}) and substituting $q\to q/\beta_\mathrm{s}$ yield
\begin{equation}
\frac{p(\beta_\mathrm{s}r)}{(\beta_\mathrm{s}r)^{D-1}}=\frac{p(r)}{r^{D-1}},
\label{eq:pexp}
\end{equation}
which is valid for $\beta_\mathrm{t}\beta^{m-2}_\mathrm{s}l_0 \lesssim r \lesssim l_0$. By Eq.~(\ref{eq:gandp}), the analogous relation takes place for the pair distribution function
\begin{equation}
\frac{g(\beta_\mathrm{s}r)}{(\beta_\mathrm{s}r)^{D-3}}=\frac{g(r)}{r^{D-3}}.
\label{eq:grexp}
\end{equation}
By using Eq.~(\ref{nrsq}) in the same manner as Eq.~(\ref{eq:prsqav}), we arrive at the relation
\begin{equation}
\frac{N(\beta_\mathrm{s}r)}{(\beta_\mathrm{s}r)^{D}}=\frac{N(r)}{r^{D}}. \label{eq:nrexp}
\end{equation}
The last three equations are equivalent.

The real-space correlation functions of fractal measure are well-studied in the 
literature, see the textbooks \cite{mandelbrot82,gouyet96:book} and references therein. 
In particular, the log-periodicity of the reduced mass-radius relation (\ref{eq:nrexp}) 
is known for ideal deterministic fractals (see, e.g., 
Refs.~\cite{badii84,bessis87,mantica07,*mantica07a}). In this paper, the mass-radius 
relation is connected to \emph{the log-periodicity} (\ref{iqlog}) \emph{in the momentum 
space for finite fractal iterations}. Conversely, once Eq.~(\ref{eq:pexp}) or 
(\ref{eq:grexp}) or (\ref{eq:nrexp}) is fulfilled, then the log-periodicity $S(q)q^D$, 
given by Eq.~(\ref{sqaffin}), is also satisfied. For a fractal embedded into the 3D 
space, the quantity $N(r)$ is \emph{radially symmetric} by definition. This means that 
the log-periodicity of $S(q)q^D$, which follows from Eqs.~(\ref{nrsq}) and 
(\ref{eq:nrexp}), is met only for the structure factor (\ref{eq:calcsf}), which includes 
the averaging over all directions of momentum. The squared anisotropic scattering 
amplitude $\prod_{i=1}^{m}|G_{i}(\bm{q})|^{2}$ does \emph{not} obey in general the 
self-affinity relation (\ref{sqaffin}), as discussed in Sec.~\ref{sec:analysis}.

The log-periodicity of the mass-radius relation (\ref{eq:nrexp}) is shown in Fig.~\ref{fig:discreteprs}c. The number of periods on the logarithmic scale coincides with the fractal iteration number.

Polydispersity makes the above log-periodicity even more apparent. Increasing the value of the distribution width $\sigma_{r}$ leads to smoothing of $p(r)$, $g(r)$ and $N(r)$. One can see a little shift of the minima to the right. From the general considerations, this is allowed as long as the distribution width is much greater than the relative shift $\Delta r/r$. The simple power law [$p(r)\varpropto r^{D-1}$, $g(r)\varpropto r^{D-3}$, and $N(r)\varpropto r^{D}$], typical for random fractals, is restored in the limit of strongly developed polydispersity.

\section{\label{sec:conclusions}{Conclusion}}

We develop a model of deterministic fractal that enables explicit analytic solutions for 
the scattering amplitude. The model generalizes the regular 3D Vicsek fractals. The 
properties of GSSVF are studied and analyzed in the both momentum and real spaces. The 
fractal dimension is controlled by the scaling parameter and can vary from $0$ to 
$2.862\ldots$. Based on the developed model, we derive analytical expressions for the 
main properties in the monodisperse and polydisperse cases: the formfactor and fractal structure 
factor, the pair distribution function, the pair distance distribution function, radius 
of gyration, intensity in zero angle and the edges of fractal region in the momentum and 
real spaces. We found the logarithmic periodicity of the intensity scaled $q^D$ times 
within the fractal region [see Eq.~(\ref{iqlog})] and relate it to the log-periodicity 
of the functions (\ref{eq:pexp}), (\ref{eq:grexp}), and (\ref{eq:nrexp}) within the 
fractal region of real space. The period is governed by the scaling parameter 
$\beta_\mathrm{s}$ and equal to $\log_{10}(1/\beta_\mathrm{s})$ on the logarithmic 
scale. This behavior of the scattering curves is explained and interpreted by analogy 
with optics (see Sec.~\ref{fractal}).

An advantage of the exact analytic solutions for deterministic fractals is the absence of phenomenological parameters, often used in the random fractal models. The behavior of random fractals can be simulated by introducing strong polydispersity in the deterministic fractal models.

The results, obtained for the GSSVF, illustrate a number of general features, common for \emph{deterministic mass fractals with a single scale}, and can be used for interpreting the experimental data. For this kind of fractal, one can extract a number of parameters from the scattering intensity, see Fig.~\ref{fig:polydispersity}:\\
\emph{(i)}  The fractal dimension from the generalized power law. \\
\emph{(ii)} The fractal scaling parameter from the period on the logarithmic scale. \\
\emph{(iii)} The number of fractal iteration, which is equal to the number of periods of function $I(q)q^D$.\\
\emph{(iv)} The lower and upper fractal edges from the diagram $I(q)q^D$ as the beginning and end of ``periodicity region". They allow us to estimate the fractal size and the smallest distance between fractal units in accordance with Eq.~(\ref{fractalrangegeneral}).\\
\emph{(v)} The total number of structural units, from which the fractal is composed, by the relation $N_m=(1/\beta_\mathrm{s})^{m D}$.\\
Note that the number of structural units can also be estimated from the ratio $I(0)/I(q_\mathrm{max})$, where $q_\mathrm{max}$ is the upper edge.

Once the internal fractal structure is known in more detail, one can derive and use 
an analytical expression for the scattering amplitude such as 
Eq.~(\ref{eq:finalformfactor}) and use it directly to fit the scattering data. This 
scheme could be applied even more efficiently to anisotropic scattering with a position 
sensitive detector.

The results obtained can be applied for various structures, whose geometries are based on iterations of fractal systems. This includes magnetic cluster structures, artificially created chemical compounds and so on.

We consider mass fractals composed of the same units, but the developed scheme allows a 
generalization to mass fractals containing units of various shapes and sizes and to 
surface fractals as well. Generalizations of the developed scheme are under way.

\begin{acknowledgments}
The authors acknowledge financial support from Russian state contract No. 02.740.11.0542 and the JINR -- IFIN-HH projects.
\end{acknowledgments}

\appendix
\section{\label{sec:hd}Hausdorff dimension}

The \emph{Hausdorff dimension} \cite{hausdorff18} can be rigorously defined as follows (see, e.g., Refs.~ \cite{rogers70:book,mandelbrot82,gouyet96:book}).

Let $A$ be a subset of the $n$-dimensional Euclidean space and $\{V_i\}$ be a covering of $A$ with $a_i=\mathrm{diam}(V_i)\leqslant a$. Then the $\alpha$-dimensional Hausdorff measure $m^\alpha(A)$ of the set $A$ is
\begin{equation}\label{hausmeasure}
m^\alpha(A)\equiv\lim_{a\to 0}\inf_{\{V_i\}}\sum_i a_i^\alpha, \quad \alpha>0.
\end{equation}
Here the infimum is on all possible covering. Note that $m^\alpha(A)$ can be infinite and $\alpha$ is not integer in general.

We define the Hausdorff dimension $D$ of the set $A$ by
\begin{equation}\label{hausdim}
D\equiv\inf\{\alpha: m^\alpha(A) = 0\}= \sup\{\alpha: m^\alpha(A) = +\infty\}.
\end{equation}
In other words, the Hausdorff dimension is the value of $\alpha$ for which the Hausdorff measure jumps from zero to infinity. For the value $\alpha = D$, this measure may be anywhere between zero and infinity.

In practice, the rigorous definition (\ref{hausdim}) is rather difficult to apply, and one can use other methods for calculating the Hausdorff  (fractal) dimension of a fractal \cite{mandelbrot82,gouyet96:book}. For instance, one can use the mass-radius relation, that is, the ``mass" of the structure
within a ball of dimension $n$ and radius $r$ centered on the fractal
\begin{equation}\label{massradius}
M(r)=A(r)r^D,
\end{equation}
where $\log A(r)/\log r\to 0$ as $r\to\infty$. By mass we mean the total fractal measure, which could be a mass, volume, surface area or any other scalar quantity attached to the fractal support.

Let us consider a few examples. For a self-similar deterministic fractal of total length $L$, whose first iteration consists of $k$ elements of size $\beta_\mathrm{s} L$, one can write $M(L)=k M(\beta_\mathrm{s}L)$. Using Eq.~(\ref{massradius}), we obtain
\begin{equation}\label{singD}
k\beta_\mathrm{s}^D=1.
\end{equation}
The Vicsek fractal, considered in Sec.~\ref{sec:construction}, is a particular case of the above fractal, and the formula (\ref{massfractaldimension}) for its fractal dimension results from Eq.~(\ref{singD}) at $k=9$. For a \emph{multiscale fractal}, giving at each iteration $k_i$ elements of size $\beta_{\mathrm{s}i} L$, we obtain
\begin{equation}\label{multD}
\sum_i k_i\beta_{\mathrm{s}i}^D=1.
\end{equation}

A more complicated example is a dense circle packing. It is an infinite set of 
non-overlapping circles of smaller and smaller radii inscribed into a larger circle in 
order to fill it completely. The set of circles is obviously a fractal. The distribution 
$n(r)$ of the circle radii is given by a simple power law $n(r)\propto r^{-\tau}$ with 
$2<\tau<3$ (see, e.g., Ref.~\cite{herrmann90,*bullett92}). The exponent $\tau$ can 
easily be related to the fractal dimension. An analog of finite iteration is the cutoff 
length $a$, for which only the circles of radii larger than $a$ are considered. The 
minimal number of disks of radius $a$ needed to cover a \emph{circle} of radius $r$ is 
proportional to $r/a$. Then the minimal number of disks for covering the fractal with a 
finite cutoff length $a$ is
\begin{equation}\label{neps}
N(a)\propto \int_{a}^{\infty}\d r\,n(r)r/a \propto a^{1-\tau}.
\end{equation}
Comparing this equation with the definition of Hausdorff dimension $N(a)\propto a^{-D}$ 
yields $D=\tau-1$. The value of dimension depends on a specific type of packing and can 
be found numerically. For instance, the value $D=1.307\ldots$ is obtained for the 
classical Apollonian packing \cite{herrmann90,*bullett92}. Note that if we consider the 
set of the \emph{filled} circles (disks), then its Hausdorff dimension is obviously 
equal to $D=2$. Indeed, the minimal number of disks of radius $a$ needed to cover a 
\emph{disk} of radius $r$ is proportional to $r^2/a^2$, and we obtain in the same manner
\begin{equation}\label{nepsdisk}
N(a)\propto \int_{a}^{\infty}\d r\,n(r)r^2/a^2 \propto a^{-2},
\end{equation}
because the integral of $n(r)r^2$ converges when $a\to0$ for $\tau<3$.

\bibliography{sasgvf}

\begin{thebibliography}{59}
\bibitem [{def()}]{defgr}%
  \BibitemOpen
  \href@noop {} {}\bibinfo {note} {Sometimes it is also called radial
  distribution function or pair correlation function.}\BibitemShut {Stop}%
\end{thebibliography}

\end{document}